\documentclass[aps,prl,twocolumn,superscriptaddress,footinbib,longbibliography,amsmath,amssymb]{revtex4-2} 

\usepackage{graphicx}
\usepackage{subfigure}
\usepackage{epsfig} 
\usepackage{dcolumn}
\usepackage{bm}
\usepackage{times} 
\usepackage{amsmath}
\usepackage{float}
\usepackage{color}
\usepackage{multirow}
\usepackage[normalem]{ulem}

\begin{document}
\title{Non-Dispersive Space-Time Wave Packets Propagating in Dispersive Media}

\author{Hao He}
\affiliation{Key Laboratory of Artificial Micro- and Nano-structures of Ministry of Education and School of Physics and Technology, Wuhan University, Wuhan 430072, China}
\affiliation{Hongyi Honor school, Wuhan University, Wuhan 430072, China}
\author{Cheng Guo}
\email{guocheng@stanford.edu}
\affiliation{Department of Applied Physics, Stanford University, California 94305, USA}
\author{Meng Xiao}
\email{phmxiao@whu.edu.cn}
\affiliation{Key Laboratory of Artificial Micro- and Nano-structures of Ministry of Education and School of Physics and Technology, Wuhan University, Wuhan 430072, China}

\bigskip

\begin{abstract}
    Space-time wave packets can propagate invariantly in free space with arbitrary group velocity thanks to the spatio-temporal correlation. Here it is proved that the space-time wave packets are stable in dispersive media as well and free from the spread in time caused by material dispersion. Furthermore, the law of anomalous refraction for space-time wave packets is generalized to the weakly dispersive situation. These results reveal new potential of space-time wave packets for the  applications in real dispersive media.
\end{abstract}

\maketitle

\section{Introduction}
Most materials are dispersive media in which waves of different  frequencies travel at different velocities.\cite{dionne2005planar,dionne2006plasmon} Dispersion causes the shape of a wave pulse to change as it travels, which is known as group velocity dispersion.\cite{saleh2019fundamentals} The group velocity dispersion limits many applications such as broadband optical communication and nanophotonics devices.\cite{edwards2008experimental,lalanne2009microscopic} In addition to dispersion, diffraction also causes the distortion of a pulse during the propagation. There have been a number of studies on diffraction-free beams, such as the well-known Bessel beams, Airy beams and X-shaped waves.\cite{durnin_diffraction-free_1987,mcgloin2005bessel,PhysRevLett.99.213901,lu1992nondiffracting,saari_evidence_1997} However, little efforts have been devoted to resist the dispersive propagation in real linear material, and the dispersion is considered as an intrinsic and inevitable nature of material.

We are interested here in resisting the temporal spread caused by dispersion of material. Achieving this goal can have tremendous implications for optical communication, where information may get lost as pulses spread in time and merge due to dispersion effect.\cite{agrawal2000nonlinear,pal2010guided} One typical strategy for material dispersion compensation in optical fiber is to strike a balance between the material dispersion and waveguide dispersion.\cite{pal2010guided} However, such compensation only works at fixed wavelength, i.e. the zero-dispersion wavelength for single-mode optical fiber.

Here we prove that a recently proposed propagation-invariant wave called space-time wave packet does not suffer from material dispersion.\cite{kondakci2019optical} Without any requirement for the medium, space-time wave packet serves as a general method for dispersion-free propagation in dispersive media. Moreover, we find that other unconventional properties of the space-time wave packets also exist in dispersive medium, for example, arbitrary group velocity.\cite{kondakci2019optical} We also modify the law of refraction for space-time wave packets, with additional terms introduced by material dispersion.\cite{bhaduri_anomalous_2020} Our work here provides a general theoretical framework for space-time wave packets in dispersive media.

The rest of the paper is organized as follows. In Sec.2~\ref{sec:review}, we validate the non-dispersive propagation of space-time wave packets in dispersive media. In Sec.3~\ref{sec:refraction}, we  modify the refraction law for space-time wave packets at the interface of two dispersive media. In Sec.4~\ref{sec:high-order}, we consider the high order contribution to the rigid propagation of space-time wave packets. We conclude in Sec.5~\ref{sec:conclusion}.

\section{Space-Time Wave Packets in Dispersive Media}\label{sec:review}

In this section, we first show the existence of space-time wave packets in dispersive media both theoretically and numerically, then consider a more specific case of quasi-monochromatic and paraxial condition in weakly dispersive media.

Consider a light pulse with scalar field $U(x,y,z,t)=A(x,y,z,t){\rm exp}[-i(\omega_0 t-n_0k_0 z)]$ propagating along the z-axis in a dispersive medium whose refraction index is frequency dependent $n=n_{(\Omega)}$, where $A(x,y,z,t)$ is the slowly varying envelope, $\omega_0$ is the frequency of the carrier, $n_0$ is the refraction index of the center frequency $\omega_0$, $k_0=\omega_0/c$ is the wave number and $\Omega=\omega-\omega_0$ is the frequency with respect to $\omega_0$.  The wave can be decomposed into plane waves:
\begin{equation}
    \begin{aligned}
        A(x,y,z;t)&=\iiint \tilde{A}(k_x,k_y,\Omega)\times\\
        &e^{i\{k_xx+k_yy-\Omega t+[k_{z(k_x,k_y,\Omega)}-n_0k_0]z\}}dk_xdk_yd\Omega
    \end{aligned}
\end{equation}
where $k_x$ and $k_y$ are components of the transverse wave vector $\bm{k}_\bot=(k_x,k_y)$ and
\begin{equation}
    k_{z(k_x,k_y,\Omega)}=\sqrt{[n(\omega_0+\Omega)/c]^2-(k_x^2+k_y^2)}
\end{equation}

Ideally, space-time wave packet consists of precisely selected plane waves such that its spectrum exhibits a special delta function:
\begin{equation}
    \tilde{A}(k_x,k_y,\Omega)=\tilde{A}_0(k_x,k_y,\Omega)\cdot \delta [\Omega-v_g(k_z-n_0k_0)]
\end{equation}
where $\tilde{A}_0(k_x,k_y,\Omega)$ is a spectrum of an arbitrary wave pulse. Substitute Equation (3) into Equation (1) and we obtain 
\begin{equation}
    \begin{aligned}
    A(x,y,z;t)&=\iint \tilde{A}_0[k_x,k_y,\Omega_{(k_\bot)}]\times\\
    & e^{i[k_xx+k_yy+(k_z-n_0k_0)\cdot (z-v_gt)]}dk_xdk_y\\
    &=A(x,y,0;t-z/v_g)
    \end{aligned}
\end{equation}
Equation (4) implies that the space-time wave packet can also exist in dispersive medium. This leads to several unconventional properties of space-time wave packet. First, the transverse field profile is unchanged during the propagation. Second, the temporal profile does not spread despite the dispersion of material. Third, the wave packet travels at a fixed group velocity $v_g$ whatever the actual dispersion is like.

\begin{figure*}
\centering
\includegraphics[width=\textwidth]{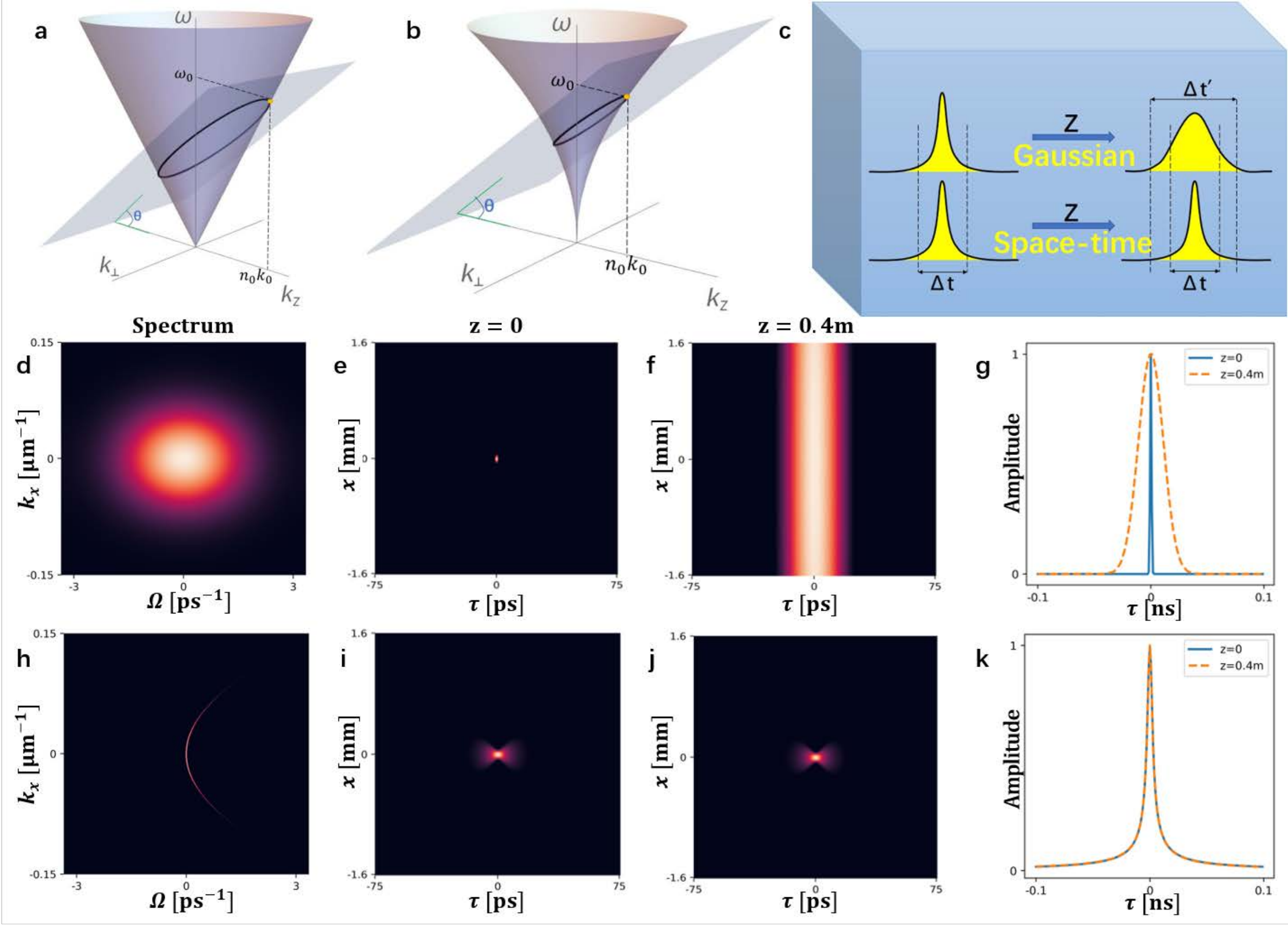}
\caption{a) Intersection of the light cone and a tilted space-time hyperplane. b) Intersection of a dispersive hypersurface and a tilted space-time hyperplane. c) Dispersive propagation of ordinary Gaussian pulse and non-dispersive propagation of space-time wave packet. d)-k) Numerical demonstration of the propagation in  dispersive medium. d) The spectrum of Gaussian pulse. e)\& f) The normalized amplitude distribution of Gaussian pulse at origin z=0 and z=0.4 m. Depicted in the moving reference frame: $\tau=t-z/v_{gauss}$, where $v_{gauss}$ is the velocity of the Gaussian pulse. Gaussian pulse suffers from both dispersion and diffraction. g) Normalized center intensity distribution of Gaussian pulse. h) The spectrum of space-time wave packet. i)\& j) The nomalized amplitude distribution of space-time wave packet at origin z=0 and z=0.4 m. Depicted in the moving reference frame: $\tau=t-z/v_{g}$. k) Normalized center intensity distribution of space-time wave packet. Space-time wave packet propagates with a fixed profile.}\label{fig:1}
\end{figure*}

To be more specific, the role of such a $\delta$ spectrum in Equation (3) can be interpreted in the following geometry. The band structure of light in free space is a 3D light cone in the 4D $(\omega,\bm{k})$ space (\textbf{Figure~\ref{fig:1}}a), and in dispersive medium the band structure is a 3D hypersurface, for example \textbf{Figure~\ref{fig:1}}b. Ordinary pulses has a 3D spectrum distribution on the 3D hypersurface because of the independence between space-time coordinates. However, the $\delta$ function, denoted by the tilted hyperplane $\Omega=v_g(k_z-n_0k_0)$ in \textbf{Figure~\ref{fig:1}}a\&b, selects a 2D distribution out of the 3D distribution, by correlating time and space. In this way one can construct a both diffraction-free and dispersion-free wave packet in dispersive media by linear superposition of the plane waves at the intersection of the hyperplane and band dispersion.

Through a rigid numerical demonstration, we validate the non-dispersive propagation of space-time wave packets, shown schematically in \textbf{Figure~\ref{fig:1}}c. Here we consider a homogeneous material with a frequency-dependent refraction index $n_{(\Omega)}=n_0+\beta\Omega$, where $n_0=0.7$, $\beta=5\times 10^{-15}$ s. A Gaussian wave packet propagates inside such a medium with its central wave length $\lambda_0=1\,{\rm \mu m}$, waist radius $W_0=30\,{\rm \mu m}$ and temporal width $\tau_0\approx1.7$ ps. Here the field is assumed to be uniform along the transverse dimension y for simplicity. \textbf{Figure~\ref{fig:1}}d shows the spectrum of the Gaussian wave packet, and \textbf{Figure~\ref{fig:1}}e\&f show the amplitude distribution at z=0 and z=0.4 m respectively (depicted in the moving reference frame). Such a Gaussian wave packet spreads in time and space as expected (\textbf{Figure~\ref{fig:1}}e\&f). And \textbf{Figure~\ref{fig:1}}g shows the nomalized amplitude of the Gaussian wave packet on the axis, which broadens due to material dispersion. By imposing the spatio-temporal correlation upon the Gaussian wave packet, we construct a space-time wave packet with group velocity $v_g=0.097c$, whose spetrum exhibits a parabolic shape(\textbf{Figure~\ref{fig:1}}h). \textbf{Figure~\ref{fig:1}}i\&j show its amplitude distribution at z=0 and z=0.4 m, and its profile hardly changed, which is in line with Equation (4). As shown in \textbf{Figure~\ref{fig:1}}k, the temporal profile of the space-time wave packet does not change despite the material dispersion. That reveals its unique resistance to material dispersion.

The resistance of space-time wave packets to temporal dispersion could be interpreted as follows: the spatio-temporal correlation forces light of different frequencies, thus different group velocities, to travel in distinct directions, resulting in the same group velocity components along the z-direction. Dispersion of different media just changes the phase velocity of every monochromatic light, and the propagation invariance of space-time wave packets remains as long as such monochromatic lights travel in their specific directions. Note here, though we formulate with a scalar field, the extenstion to a vector field is straight forward. 

However, such precise control over $\Omega$ and $\bm{k}_\bot$ is impossible in practice due to the infinite intensity such $\delta$ function requires and the finite precision in experiments.\cite{porras_nature_2018,murat_yessenov_what_2019} In addition, the numerical aperture is also restricted in a small range in most cases.\cite{hall_spectral_2020} So it stands critical to explore the more realistic case, with narrow bandwidth, limited control over the spectrum and paraxial condition. For a space-time wave packet which is quasi-monochromatic, the frequency-dependent refraction index can be expanded in the vicinity of its center frequency $\omega_0$ as $n_{(\Omega)}\approx n_0+\beta \Omega$, and the spectrum reduces to the parabolic form:
\begin{equation}
    \Omega\approx\frac{c}{2\alpha n_0\omega_0}k_\bot^2
\end{equation}
where $\alpha=(n_0+\beta\omega_0)/c-1/v_g$. So the parabolic spectrum shown in Equation (5) suffices to sculpt the spatio-temporal spectrum depicted in \textbf{Figure~\ref{fig:1}}h for a dispersive medium, as long as the omitted high order terms in the dispersion relation are relatively small compared with $\beta\Omega$,

\section{Refraction of Space-Time Wave Packets at Interface of Dispersive Media}\label{sec:refraction}
One unique property of space-time wave packets is the anomalous refraction at the interface of two non-dispersive media, where some corollaries of the Snell's law no longer hold.\cite{bhaduri_anomalous_2020} Here we extend the study to the interface of two possibly dispersive media, and modify the law of refraction for space-time wave packets.

When travelling through different media of distinct dispersion relation, the spatio-temporal correlation is expected to be conserved, because such correlation exhibits the same form of a parabola shown in Equation (5) in different weakly dispersive media. To be more specific, with respect to the normal incidence of space-time wave packets between two dispersive media, conservation of energy and transverse momentum lead to the invariance of $\omega$ and $\bm{k}_\bot$ for every monochromatic component. Therefore the parabolic spectrum in Equation (5) is also preserved under the refraction, so is the value of the coefficient. Such invariant coefficient, expressed in two dispersive media separately, gives rise to the refraction law for space-time wave packets.
\begin{equation}
    n_{01}(n_{01}+\beta_{1}\omega_0-\tilde{n}_1)=n_{02}(n_{02}+\beta_{2}\omega_0-\tilde{n}_2)
\end{equation}
where $n_{0i}$ is the refraction index of the medium $i(i=1,2)$ at the center frequency, $\tilde{n}_i=c/v_{gi}$ is the group index and $\beta_{i}$ refers to the first order of dispersion. For non-dispersive medium, the refraction law above reduces to:
\begin{equation}
    n_{01}(n_{01}-\tilde{n}_1)=n_{02}(n_{02}-\tilde{n}_2)
\end{equation}
which agrees with the result of Bhaduri et al..\cite{bhaduri_anomalous_2020}

\begin{figure*}
\centering\includegraphics[width=\linewidth]{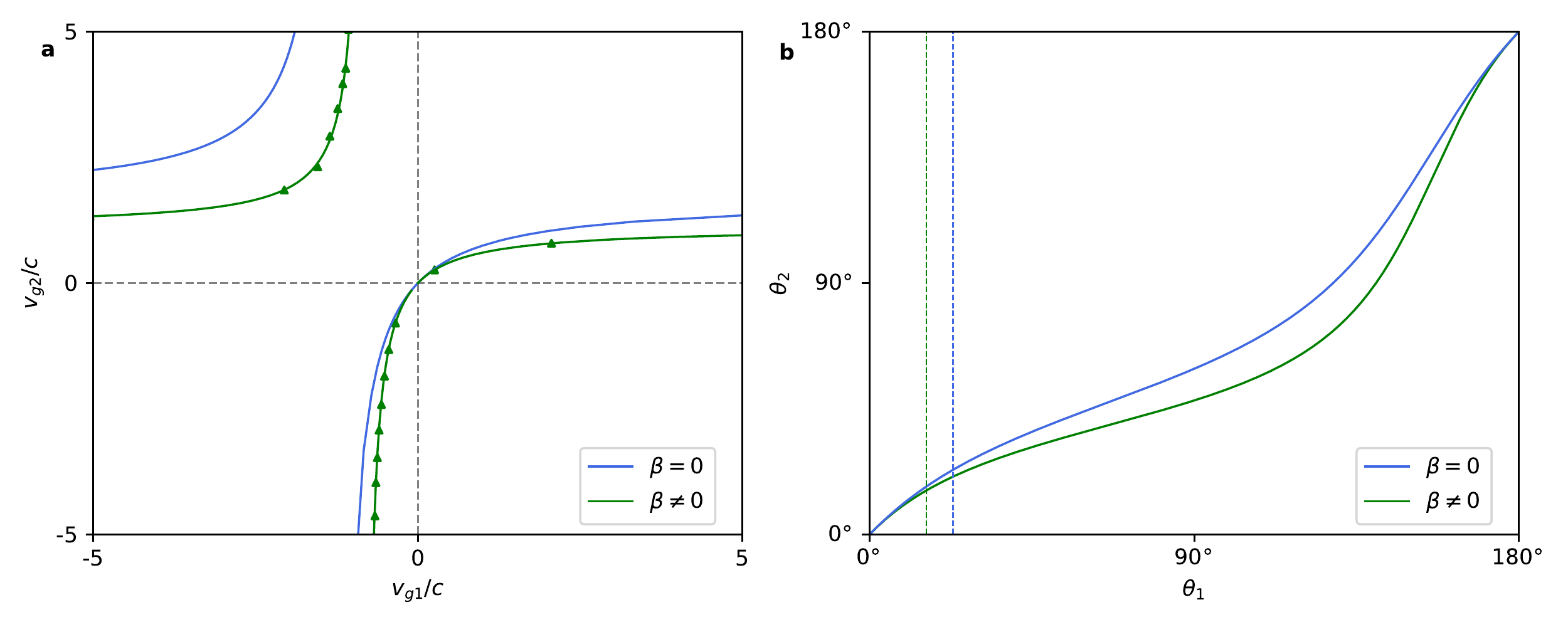}
\caption{a) The group velocity relation between the two media in the normal incidence. b) The group velocity relation expressed in terms of the tilt angle $\theta$ of the space-time hyperplane. The dashed lines in b) represent the equivalence of $\theta_1$ and $\theta_2$, i.e., $v_{g1}=v_{g2}$. The left hand side of the dashed line corresponds to $v_{g1}<v_{g2}$ (anomalous) and right hand side corresponds to $v_{g1}>v_{g2}$ (normal).}\label{fig:2}
\end{figure*}

Although the dispersion near $\omega_0$ can be relatively small: $\beta\Omega\ll n_0$, such a small term could actually cause appreciable change in the refraction of space-time wave packets, as the change introduced by $\beta$ is actually $\beta\omega_0$. To demonstrate its effect, \textbf{Figure~\ref{fig:2}} compares two refraction processes both from non-dispersive air to a weakly dispersive medium $\beta=1.64\times 10^{-16}$ s, $\omega_0=1.89\times 10^{15}$ s$^{-1}$ and $n_0=1.34$ (green line) and a non-dispersive medium $\beta=0$ with other parameters the same (blue line). In this figure, we show the law of refraction Equation (6) as a relation between $(v_{g1},v_{g2})$ and $(\theta_1,\theta_2)$, here $\theta=acot\tilde{n}$ stands for the tilt angle of space-time hyperplane shown in \textbf{Figure~\ref{fig:1}}b. Both the two lines in \textbf{Figure~\ref{fig:2}}a have the shape of the inverse proportional function [indicated by Equation (6)] and the existence of dispersion just shifts the center of such shape and stretches the coordinates. The green triangles in \textbf{Figure~\ref{fig:2}}a represent the numerical results of the refraction process making use of Fresnel equation, which validates Equation (6). In terms of the tilt angle $\theta$, the normal dispersion ($\beta>0$) tends to ``drag" the blue line in \textbf{Figure~\ref{fig:2}}b downward. Meanwhile, compared with ordinary pulsed beams with only fixed group velocity, the group velocity of space-time wave packets can increase, decrease or remain the same in the refraction process depending on the group velocity of incident space-time wave packets [\textbf{Figure~\ref{fig:2}}b]. The region to the left of the dashed lines (Blue and green denote respectively, the nondispersive and dispersive cases.) in \textbf{Figure~\ref{fig:2}}b corresponding to $v_{g1}<v_{g2}$ stands for the ``anomalous" refraction of space-time wave packets, because the group velocity does not decrease as expected when $n_1<n_2$. Compare the blue and the green dashed lines, we can conclude from \textbf{Figure~\ref{fig:2}}b that the dispersion here also can change the range of group velocity in ``anomalous" refraction.

\begin{figure*}
\centering\includegraphics[width=\linewidth]{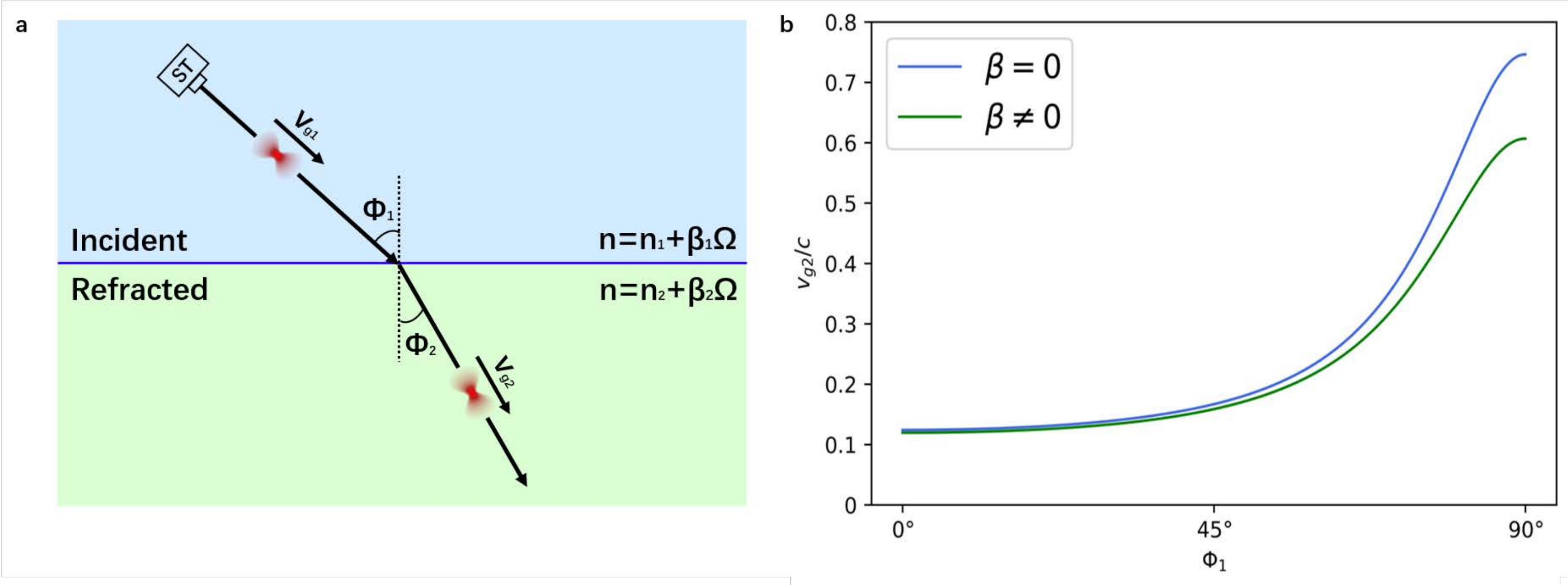}
\caption{a) Inclined refraction of space-time wave packet. b) The relation between incident angle $\Phi_1$ and the group velocity $v_g$ of refracted wave. }\label{fig:3}
\end{figure*}

For oblique incidence (\textbf{Figure~\ref{fig:3}}), the condition $k_{x1}=k_{x2}$ turns into $k_{x1}cos\Phi_1=k_{x2}cos\Phi_2$ as $k_x$ is no longer parallel to the interface.\cite{bhaduri_anomalous_2020} The law of refraction for space-time wave packet changes correspondingly:
\begin{equation}
    n_{01}(n_{01}+\beta_{1}\omega_0-\tilde{n}_1)cos^2\Phi_1=n_{02}(n_{02}+\beta_{2}\omega_0-\tilde{n}_2)cos^2\Phi_2,
\end{equation}
where $\Phi_1$, $\Phi_2$ are the angle of incidence and refraction respectively, determined by Snell's law: $n_1sin\Phi_1=n_2sin\Phi_2$. For instance, an incident space-time wave packet with group velocity $v_g=0.1c$ could be refracted to have a new group velocity in a wide and continuous range when it comes from different $\Phi_1$, as illustrated in \textbf{Figure~\ref{fig:3}}b, and the existence of dispersion tends to change such velocity range from approximately 0.1c$\sim$ 0.75c to 0.1c$\sim$ 0.6c. Therefore, the dispersion here narrows (can also broaden in case of anomalous dispersion) the $v_g$ range of the refracted wave.

\section{High Order Terms in the Spectrum}\label{sec:high-order}
In practice, sculpting a spectrum of accurate parabola shape is sometimes intractable, \cite{bhaduri_anomalous_2020,guocheng} but a polynomial spectrum also provides additional measure to shape the propagation.\cite{yessenov_engineering_2021} A more general case is to reconsider the relatively small high order terms in the parabolic spectrum, for example:
\begin{equation}
    \Omega(\boldsymbol{k}_\bot)\approx\alpha_1(k_x^2+k_y^2)+\alpha_2(k_x^2+k_y^2)^2
\end{equation}
For simplicity, we discuss its behavior in free space. And the result will deviate slightly from the case for ideal space-time wave packets:
\begin{equation}
    \begin{aligned}
        A(x,y&,z;t)\propto\iint \tilde{A}e^{i(k_xx+k_yy)}\times\\
        &{\rm exp}[-i\alpha_1 k_\bot^2(t-z/v_g)-i\alpha_2k_\bot^4(t-z/v')]dk_xdk_y
    \end{aligned}
\end{equation}
where$1/v_g=1/c-c/(2\omega_0\alpha_1)$ is the reciprocal of group velocity and $1/v'=1/c-c^3/(8\omega_0^3\alpha_2)$ is the high term contribution with the dimension of the inverse of speed controlled by $\alpha_2$. Therefore the previous rigid propagation fails to satisfy as the amplitude distribution has a nonvanishing z-dependence:
\begin{equation}
    \begin{aligned}
        A(x,y&,z;t)\propto\iint \tilde{A}e^{i(k_xx+k_yy)}\times\\
        &{\rm exp}[-i\alpha_1 k_\bot^2t'-i\alpha_2k_\bot^4(t'-z/v'+z/v_g)]dk_xdk_y
    \end{aligned}
\end{equation}
where $t'=t-z/v_g$. Such a deviation would  gets larger at long distance or time.

However, the paraxial condition ensures $A(x,y,z;t)\approx A(x,y,0;t-z/v_g)$ again at least within a finite space and time since $\alpha_2k_\bot^2\ll\alpha_1$, and actually the propagation distance of space-time wave packets is also finite due to the ``fuzziness" in the spatio-temporal correlation.\cite{murat_yessenov_what_2019} Such deviation, therefore, can be relatively small in the propagation of space-time wave packets. In fact, such deviation corresponds to the first dispersive term of space-time wave packets in free space.\cite{yessenov_engineering_2021} To be more specific, only one of the two terms in $v'$ is introduced by $\alpha_2$ while the other comes from the high terms in Taylor series, which is previously omitted in the expansion of $k_z$. In other words, one can make $A(x,y,z;t)= A(x,y,0;t-z/v_g)$ strictly true at the order of $k_\bot^4$ if we choose $\alpha_2$ properly such that $v'=v_g$, in which case the z-dependence in Equation (11) vanishes.

Furthermore, one can always complete Equation (9) with additional high terms correspondingly to compensate the destruction of rigid propagation that is caused by the omitted terms in the expansion of $k_z$. The more terms one compensates, the closer the spectrum is to the $\delta$ function, thus closer to the ideal space-time wave packet. From another point of view, putting more power series terms in Equation (9) can make the free space to exhibit arbitrary magnitude, sign and order of dispersion.\cite{yessenov_engineering_2021} For example, a linear term added in Equation (9) will result in an effective group-velocity dispersion (GVD), which has been demonstrated theoretically and experimentally by Murat Yessenov et al..\cite{hall2021vwaves}

\section{Discussion and Conclusion}\label{sec:conclusion}

Until now, space-time wave packets have mostly been investigated in free space. \cite{hall_spectral_2020,motz_axial_2020,yessenov_engineering_2021} However, there has been attempt using spatio-temporal correlation in surface plasmon polaritons.\cite{schepler_space-time_2020} More generally, our work presented here reveals the great potential of imposing spatio-temporal frequency correlation in the dispersive media, which is proved to resist the temporal broadening in dispersive medium. Here, the first dispersion term in the medium provides a handle to change the group velocity without destroying the spatio-temporal correlation. Meanwhile, the higher order dispersion do not affect the the propagation of space-time wave packet, although that always leads to some unique properties to other diffraction-free wave like an Airy pulse.\cite{driben2013inversion} Apart from that, we generalize the refraction law for space-time wave packet to dispersive media. These findings further support the idea of imposing spatio-temporal correlation to shape the propagation of wave, and pave the way for dispersion resistance in dispersive media.

The recent method proposed to impose precise spatio-temporal correlation with a compact photonic crystal slab freed the generation of space-time wave packets from sophisticated procedure using a two-dimensional pulse shaper.\cite{guocheng,bhaduri_anomalous_2020} On that basis, we prove that the possibly high order terms in the band dispersion of such photonic crystal slab actually open up a new avenue for sculpting the arbitrary dispersion for space-time wave packets.\cite{yessenov_engineering_2021} 

\bigskip

\begin{acknowledgments}
    This work is supported by the National Natural Science Foundation of China (Grant No.11904264).
\end{acknowledgments}

\bibliography{main}{}

\clearpage

\end{document}